\documentclass[fleqn,usenatbib]{rasti}

\usepackage{newtxtext,newtxmath}

\usepackage[T1]{fontenc}
\usepackage{ulem}
\usepackage{lipsum}
\usepackage{anyfontsize}

\DeclareRobustCommand{\VAN}[3]{#2}
\let\VANthebibliography\thebibliography
\def\thebibliography{\DeclareRobustCommand{\VAN}[3]{##3}\VANthebibliography}

\usepackage{graphicx}	
\usepackage{amsmath}	
\usepackage{xcolor} 

\newcommand{\code}[1]{\texttt{#1}}

\title[Stellar flares with Ariel]{Stellar flares with Ariel: detection prospects and impacts on exoplanet spectroscopy}

\author[K. Vida, B. Seli \& B. Edwards]{
K. Vida,$^{1,2}$\thanks{E-mail: vidakris@konkoly.hu}
B. Seli$^{1,2}$
B. Edwards$^{3,4}$
\\
$^{1}$Konkoly Observatory, HUN-REN Research Centre for Astronomy and Earth Sciences, Konkoly Thege Miklós út 15-17., H-1121, Budapest, Hungary\\
$^{2}$HUN-REN CSFK, MTA Centre of Excellence, Budapest, Konkoly Thege Miklós út 15-17., H-1121, Budapest, Hungary\\
$^{3}$Department of Physics and Astronomy, University College London, Gower Street, WC1E 6BT London, UK\\
$^{4}$SRON, Space Research Organisation Netherlands, Niels Bohrweg 4, NL-2333 CA, Leiden, The Netherlands\\
}

\date{Accepted XXX. Received YYY; in original form ZZZ}

\pubyear{2022}

\begin{document}
\label{firstpage}
\pagerange{\pageref{firstpage}--\pageref{lastpage}}
\maketitle

\begin{abstract}
Stellar flares are important tracers of magnetic activity and can significantly influence the characterization of exoplanet host stars by introducing time-dependent astrophysical variability. 
ESA's Ariel mission, designed for atmospheric characterization of exoplanets, will include the VISPhot instrument, providing high-cadence optical photometry offering an opportunity to study stellar flares while identifying and mitigating their impact on transit spectroscopy. 
We discuss Ariel's potential for flare studies, emphasizing the importance of short-cadence observations for resolving flare morphology and estimating flare energies. 
We identify potential flaring targets within the Ariel Mission Candidate Sample by cross-matching it with a TESS-based flare catalog, identifying 44 stars in the sample of 3321 host stars with flare rates spanning 0.0012--1.05 flares per day yielding a lower limit. 
The majority of these active stars are expected to flare infrequently during typical Ariel observations, but for a few of the most active targets there is a significant chance of flaring during a typical Ariel observing window. 
We further investigate the detectability of stellar flares in Ariel's wavelength range using synthetic spectra based on time-dependent \code{RADYN} flare models propagated through the Ariel radiometric simulator. 
Our simulations indicate that strong flares on cool dwarfs can produce measurable infrared signatures, with flux variations reaching several percent, substantially exceeding the amplitudes of typical exoplanet atmospheric features. 
These results demonstrate stellar flares represent both a valuable ancillary science case and a potential source of contamination for Ariel observations, highlighting the need for improved flare models and mitigation strategies to maximize the mission's scientific return.

\end{abstract}

\begin{keywords}
Data Methods -- Instrumentation -- Space Telescopes -- Stellar Flares
\end{keywords}


\section{Ariel fast photometry as a tool for flare studies }

Stellar flares are fast, energetic transient events that occur due to the reconnection of surface magnetic fields on timescales of minutes to hours. These energetic outbursts are a direct manifestation of magnetic activity on stars, providing critical insight into stellar magnetic fields, energy release mechanisms, and space weather conditions. 
Beyond their stellar-physics implications, flares are also highly relevant for exoplanetary environments, as repeated high-energy irradiation may alter atmospheric chemistry, erode planetary atmospheres, and influence long-term habitability \citep{2007AsBio...7..167K, 2008SSRv..139..437Y}.

Because flares evolve rapidly, accurate characterization critically depends on the temporal resolution of the observations. Insufficient cadence can significantly distort inferred flare properties such as peak amplitude, duration, morphology, and total emitted energy \citep{flatwrm}. In particular, coarse sampling may smooth over complex multi-peaked structures and systematically underestimate flare energies. At the same time, flares occur stochastically and cannot be predicted in advance, making them both scientifically valuable transient events and a potential source of "astrophysical noise" for exoplanet characterization missions such as Ariel \citep{tinetti2018}.

\begin{figure}
    \centering
    \includegraphics[width=1\columnwidth]{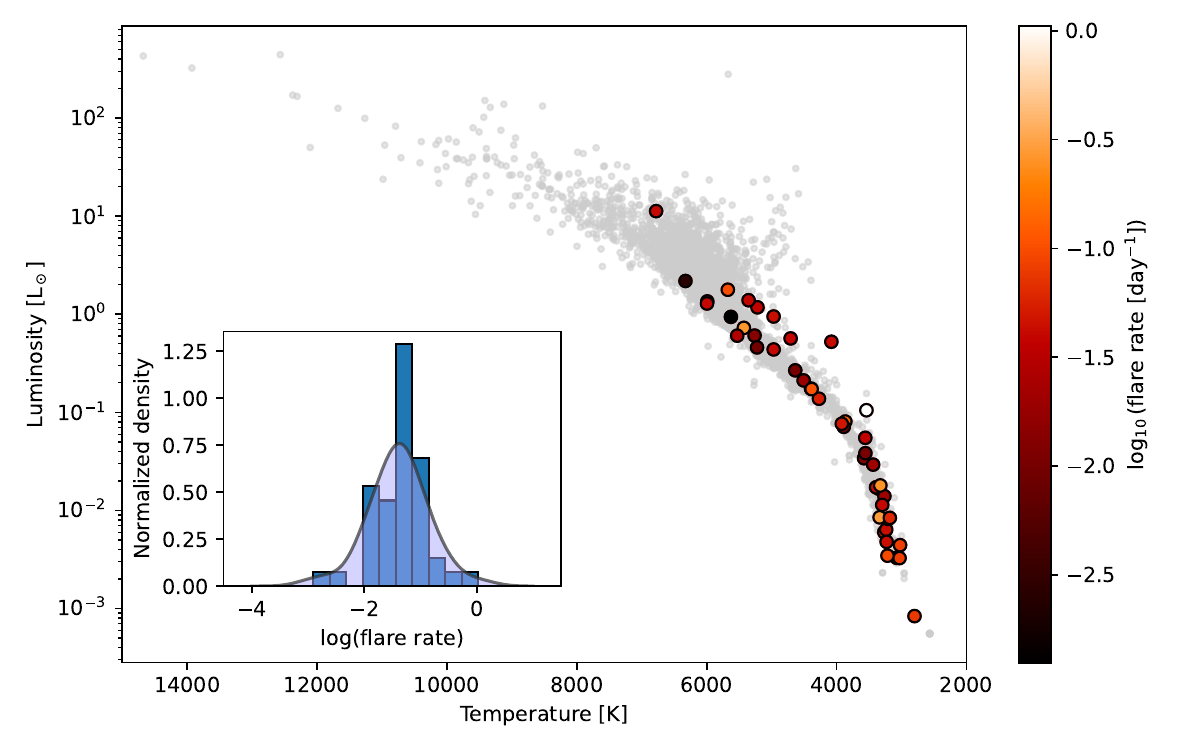}
    \caption{Hertzsprung–Russell diagram of the Ariel {planet entries of the} target list \citep{ArielTargetList}  {(see text for details)}
    with flaring stars highlighted (color scaled with flaring rate {in flares/day}). The inset shows a histogram and kernel density estimation of the flare rates.}
    \label{fig:hrd}
\end{figure}

Besides spectroscopic instruments, Ariel will also carry VISPhot, 
an instrument that will provide fast optical photometry, enabling high-cadence monitoring of transient events with much better time resolution than is usually available in survey data, reaching  10 Hz sampling rate in the optical regime (between 500 and 600\,nm). 
High-resolution photometry plays a vital role in the study of fast stellar transients: accurate flare parameter estimation, such as energy, depends critically on the time sampling of the light curve. Coarse temporal resolution---like Kepler's long cadence---can obscure complex flare structures and lead to energy underestimation \citep{2018ApJ...859...87Y}. 
VISPhot's unprecedented frequency and precision will not only be useful to detect exoplanet transits, it will also allow us to obtain much more precise energy estimations for stellar flares. 


Except for a few early, experimental observations \citep[e.g., ][]{1988SvAL...14...65B}, flares were not routinely observed with fast photometry, leaving a gap in our ability to accurately characterize them. This changed with efforts like the one at the Piszkéstető Observatory \citep{Schmercz2026A&A...708A.225S},  utilizing a 1-meter telescope equipped with the OCELOT EMCCD camera, capable of sub-second exposures. 
\cite{Schmercz2026A&A...708A.225S} detected 42 flares in 211 h of 0.3-s $B$-band monitoring of AD Leo. Two events showed candidate quasi-periodic pulsations with periods of approximately 1 and 3 min, whereas no significant variability was found below a few seconds. For two of the larger double-peaked flares, morphology-based complexity remained nearly unchanged when the light curves were binned to approximately 4--5 s and declined at coarser sampling. These results suggest that the approximately 1-Hz photometry expected for Ariel core-survey products should resolve the empirically identified flare timescales, while modest 1--5 s co-addition may improve signal-to-noise without erasing the dominant morphology. Note, that this inference is not universal, as it is based on a single active M dwarf observed in the bluer $B$ band.
{However, the observations of Ariel can serve as an important observational input for flare physics that is in dire need of multi-band observations: Ariel can detect flares at high contrast in VISPhot, while 
VISPhot–FGS color evolution can  constrain the flare continuum/filling factor. We can also test whether morphology changes with wavelength, search for Paschen/Brackett/Pfund emission and continuum changes in the spectroscopic channels and use the optical/near-IR time series as priors in a joint transit--flare fit. }

{Active stars with frequent flaring {therefore} could be interesting targets for Ariel's ancillary observing list between the targets of the primary mission.  According to \cite{ancillary_time}, waiting times could add up to 19--27\% of the mission duration, and a few hours of observing time could be enough to catch flares on stars with high levels of activity, while it does not necessarily require continuous monitoring.}

\section{Characterization of possible flaring Ariel targets}

The proposed target list of Ariel \citep{ArielTargetList} focuses on host stars with low level of activity, thus it is unlikely that  most of the prime scientific targets would show flare during the observations. For validation we cross-matched the target list of Ariel (Mission Candidate Sample version 2026--05--11\footnote{\url{https://github.com/arielmission-space/Mission_Candidate_Sample}}) with the flare catalog of \cite{flarecatalog}\footnote{\url{https://zenodo.org/records/14179313}}. 
In this catalog, flaring stars were selected from the 2-minute cadence observations obtained by the Transiting Exoplanet Survey Satellite (TESS) across mission sectors 1--69. The authors analyzed the light curves using a retrained implementation of the \texttt{flatwrm2} pipeline, which employs a long short-term memory (LSTM) neural network to automatically identify flare-like transient brightenings. Before flare detection, the light curves were detrended and normalized in order to suppress long-timescale stellar variability and instrumental systematics that could mimic flare events.

Candidate flares were identified by searching for impulsive increases in flux above the local background level. The neural-network classifier was trained to distinguish astrophysical flares from artifacts such as scattered-light contamination, momentum dumps, eclipses, pulsations, and noise spikes. Additional filtering steps were then applied to reject events with poor temporal coverage, low signal quality, or morphologies inconsistent with classical stellar flares. This combination of automated classification and post-processing was intended to reduce contamination while preserving statistically reliable flare detections.

The final sample of flaring stars was produced using a conservative vetting strategy that emphasized catalog purity over completeness. After the automated selection stage, the remaining candidates underwent manual inspection to eliminate residual false positives and ambiguous cases. Only stars with clean, well-resolved flare events were retained for subsequent morphology analysis. The resulting catalog therefore contains a highly reliable set of flare stars and flare events suitable for studying how flare morphology varies across stellar spectral types and activity regimes.
The cross-match between the \cite{flarecatalog} catalog and the Ariel Mission Candidate Sample has identified 44 potential flaring Ariel targets with flare rates of 0.0012--1.0547 events per day, making them promising candidates for further study.  
For the analysis we used the latest, 2026-05-11 version of the combined list of known exoplanets and TESS Planet Candidates (TPCs) containing 3321 host stars.
The physical properties and flaring rate of these targets are summarized in Table \ref{tab:flaring_targets}. 
Note, that the flare rates derived here should be interpreted as observed catalog rates rather than completeness-corrected occurrence rates. The \cite{flarecatalog} catalog was constructed to maximize sample purity and support flare-morphology studies; its strict filtering and manual vetting reduce sensitivity to low-amplitude, short-duration, or poorly resolved events, particularly in noisier light curves. The true flare rates may therefore be higher than the reported values, and the corresponding Ariel overlap probabilities should be regarded as lower-limit estimates above the TESS detection threshold. 

In Figure \ref{fig:hrd}, we plotted the Hertzsprung--Russell diagram of the Ariel Mission Candidate Sample and the flaring targets in the sample; and the flaring rate of the targets. From these we can see that most of the flaring targets show only a low observed flare rate with the kernel density estimation peaking at 0.042 flares per day. Assuming a typical flare length of 30 minutes $\tau_\mathrm{flare}$, a typical Ariel of observing window of $\sim10$ hours $T_\mathrm{obs}$, and a Poisson-distributions of the flares; the chance of a flare {overlapping} this observing window is $<1\%$ for most all targets where the flaring rate $r$ is lower than 0.023 flares/day using
\[
P_{\rm overlap}=1-\exp[-r(T_{\rm obs}+\tau_{\rm flare})].
\]
For the most active target in the mission candidate sample (TIC~441420236 or AU~Mic) this chance is roughly 37\%. 

Recently, \cite{Galletta2026} independently presented a complementary analysis of stellar flaring among Ariel targets. They reanalysed TESS light curves for 290 nearby confirmed hosts from the July 2024 Mission Candidate Sample, deriving completeness-tested flare-frequency distributions and energy-dependent probabilities of flare contamination during individual planetary transits. In contrast, our cross-match uses the May 2026 combined list of known planets and TESS Planet Candidates and an independent, conservatively vetted flare catalogue; consequently, the rates reported here should be interpreted as lower limits rather than completeness-corrected occurrence rates. The systems common to the two studies are marked with asterisk in Table~\ref{tab:flaring_targets}. The independent identification of active systems such as AU Mic, DS Tuc, HIP 67522 and V1298 Tau supports the conclusion that a small subset of Ariel hosts requires explicit flare-aware observing and analysis strategies. Quantitative probabilities from the two studies should not be compared directly, however, because they employ different flare-selection functions, energy thresholds and observing windows. Our subsequent analysis focuses on the complementary question of how time-dependent flare spectra would appear across Ariel's photometric and spectroscopic channels.

\begin{table}
\caption{Summary of the flaring targets in the Ariel mission candidate sample showing the targets IDs in the TESS input catalog, their name, temperature, radius, the number of observed TESS sectors ($S$) and their flaring rate. 
{Targets also present in the analysis of {\protect\cite{Galletta2026}} are marked with an asterisk.} }
\label{tab:flaring_targets}
    \begin{tabular}{cccccc}
\hline
\hline
TIC ID & Name &$T_\mathrm{eff}^\dagger$ & Radius$^\dagger$& $S$ & Flare rate$^\ddagger$ \\
&&[$K$] & [$R_\odot$]& &[day$^{-1}$]\\
\hline
 441420236$^*$  &     AU Mic    &3540   &0.862  &      2        &       1.055 \\
 098796344$^*$  &  LTT1445A     &3337   &0.276  &      2        &       0.284 \\
 410214986$^*$  &    DS Tuc     &5428   &0.964  &      5        &       0.262 \\
 299798795      &  TOI-1224     &3326   &0.404  &      8        &       0.253 \\
 166145361      &  TOI-6599     &3075   &0.202  &      2        &       0.136 \\
 071268730      &  TOI-5375     &3865   &0.633  &      4        &       0.134 \\
 434226736$^*$  &     K2-25     &3207   &0.293  &      1        &       0.133 \\
 058472992      &  TOI-4625     &4384   &0.720  &      2        &       0.108 \\
 166527623$^*$  &  HIP67522     &5675   &1.380  &      3        &       0.098 \\
 200322593$^*$  &   TOI-540     &3216   &0.190  &      5        &       0.095 \\
 459837008      &  TOI-2267     &3022   &0.243  &      8        &       0.083 \\
 230741378$^*$  &SPECULOOS-3    &2800   &0.123  &      1        &       0.076 \\
 336961891      &  TOI-4642     &3266   &0.242  &      2        &       0.064 \\
 086263325$^*$  &  TOI-3884     &3180   &0.302  &      2        &       0.061 \\
 157667414      &  TOI-4576     &4271   &0.678  &      3        &       0.054 \\
 362249359$^*$  &   TOI-833     &3920   &0.600  &      6        &       0.048 \\
 015756231$^*$  &  V1298 Tau    &4970   &1.314  &      2        &       0.043 \\
 337217173      &  TOI-7049     &3236   &0.254  &      3        &       0.043 \\
 101955023$^*$  &    GJ1132     &3229   &0.221  &      1        &       0.043 \\
 056658270      &IRAS 04125+2902        &4080   &1.450  &      3        &       0.042 \\
 035582553      &   TOI-649     &6779   &2.433  &      1        &       0.041 \\
 460205581$^*$  &   TOI-837     &5995   &1.052  &      6        &       0.040 \\
 146520535$^*$  &   TOI-942     &4969   &0.893  &      1        &       0.040 \\
 435878153      &  TOI-6258     &3392   &0.380  &      1        &       0.040 \\
 368287008$^*$  &  TOI-2015     &3297   &0.327  &      1        &       0.040 \\
 315398848      &  TOI-3314     &5355   &1.370  &      1        &       0.039 \\
 217933560      &  TOI-5051     &4708   &1.132  &      1        &       0.039 \\
 238920875      &  TOI-2329     &5218   &1.328  &      3        &       0.038 \\
 161223358      &  TOI-2415     &3558   &0.617  &      1        &       0.037 \\
 257605131$^*$  &   TOI-451     &5530   &0.848  &      3        &       0.028 \\
 264301607      &  TOI-5278     &3349   &0.383  &      5        &       0.024 \\
 377293776$^*$  & TOI-1450A     &3437   &0.483  &      2        &       0.021 \\
 455947620      &  TOI-6022     &3264   &0.370  &      2        &       0.021 \\
 464646604$^*$  &  HIP 94235    &5991   &1.080  &      2        &       0.020 \\
 369789627      &  TOI-3258     &4508   &0.754  &      2        &       0.019 \\
 047937361      &  TOI-4131     &5263   &0.937  &      3        &       0.013 \\
 224298134      &  TOI-2079     &3577   &0.481  &      3        &       0.013 \\
 302798365      &  TOI-2946     &4394   &0.721  &      3        &       0.013 \\
 262530407$^*$  &    GJ3090     &3556   &0.516  &      4        &       0.011 \\
 046631742      &  TOI-5358     &4636   &0.803  &      4        &       0.010 \\
 443616612      &     K2-43     &3890   &0.587  &      4        &       0.010 \\
 027010191      &  KOI-7368     &5225   &0.827  &      4        &       0.010 \\
 149603524$^*$  &   WASP-62     &6327   &1.230  &     30        &       0.003 \\
 391903064      &  TOI-3353     &5627   &1.021  &     33        &       0.001 \\

\hline\end{tabular}

$^\dagger$ from the 2026-05-11 version of Mission Candidate Sample, based on  \cite{ArielTargetList}\\
$^\ddagger$from \cite{flarecatalog}
\end{table}
\section{The possible effects of flaring on Ariel observations}
\begin{figure*}
	\includegraphics[width=\textwidth]{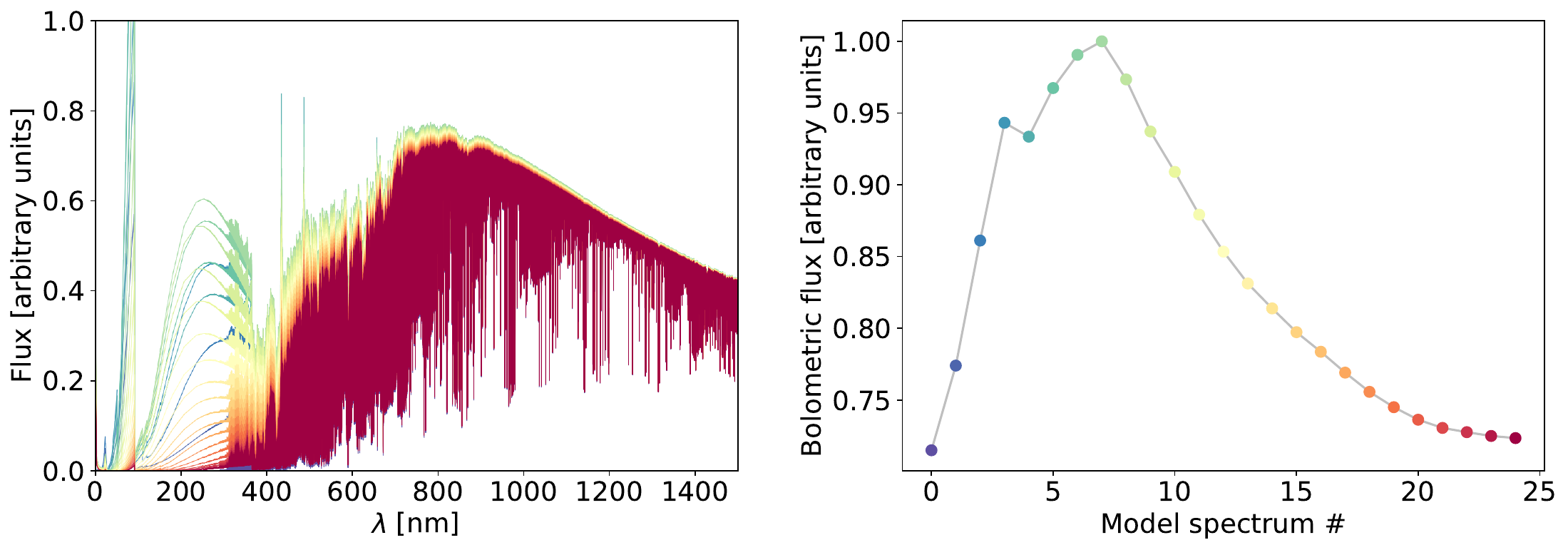}
    \caption{Contribution of a flare to the observed spectrum on anM-dwarf using a BT-Settl model spectrum \citep{2014IAUS..299..271A} and a time-series flare model using the \code{RADYN} code \citep{2024ApJ...969..121K}.}
    \label{fig:btsettl}
\end{figure*}

\begin{figure}
    \centering
    \includegraphics[width=\columnwidth]{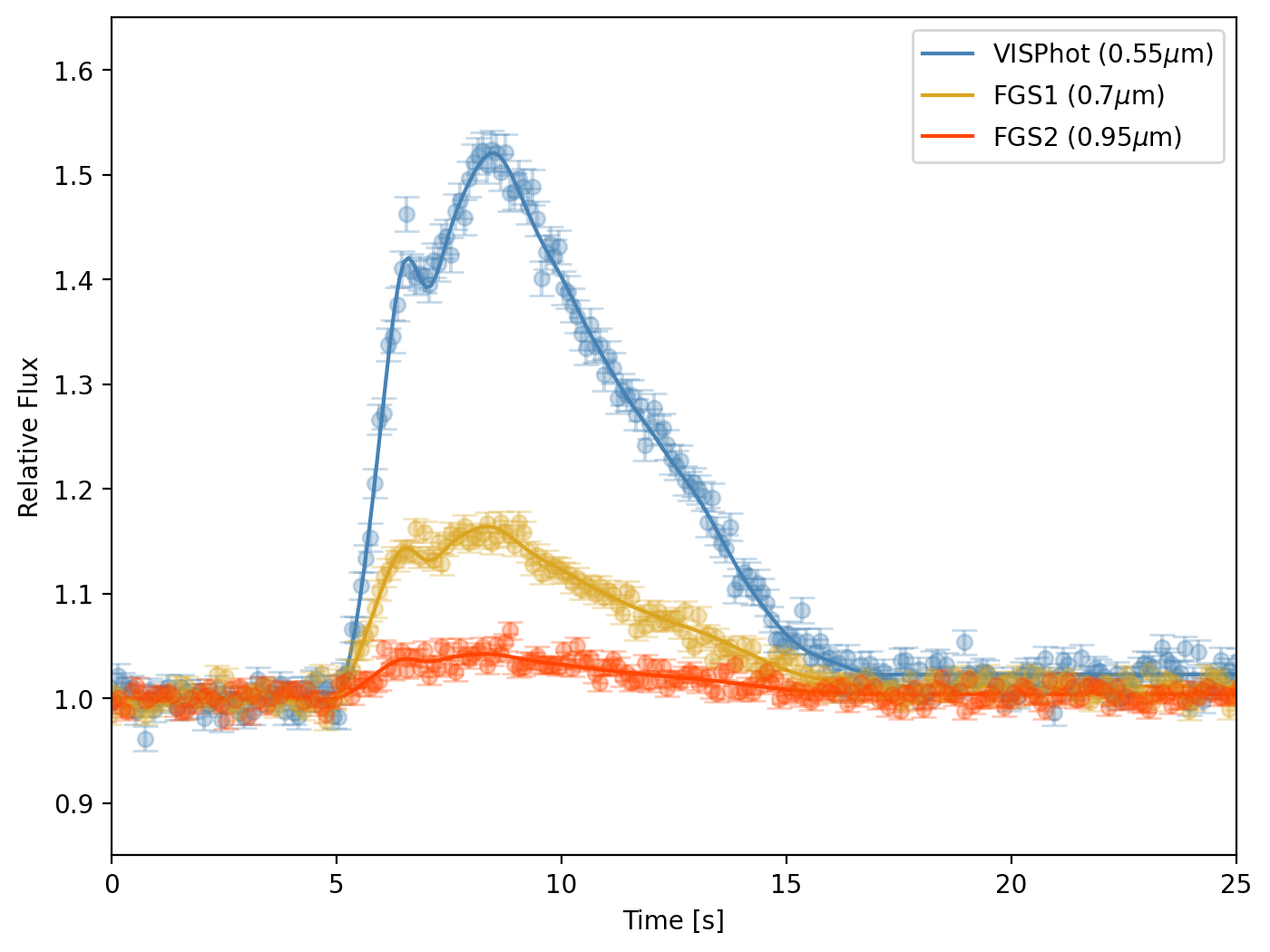}
    \includegraphics[width=\columnwidth]{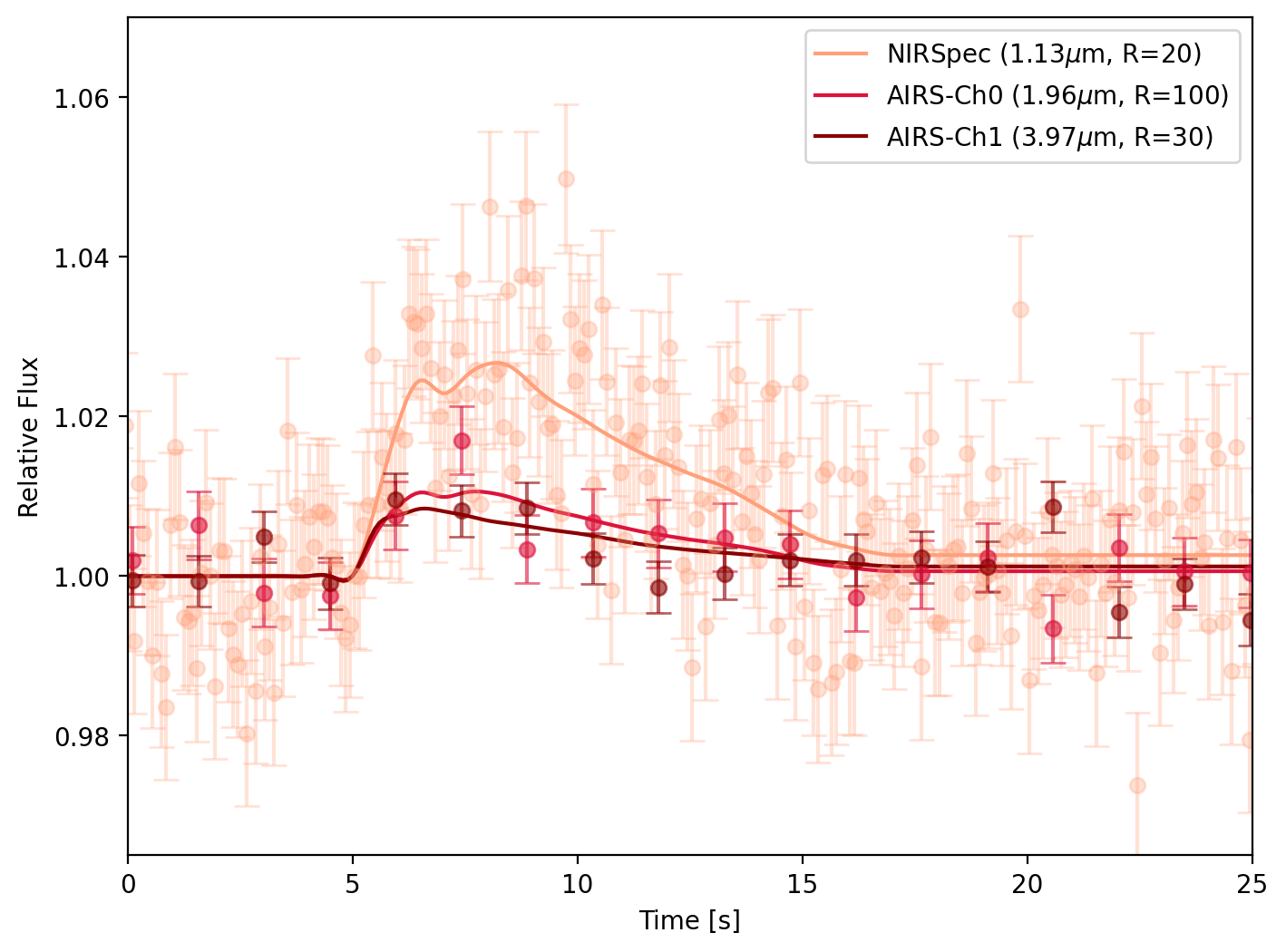}
    \caption{Simulated Ariel time-series of a flare on GJ367. As expected, the flare is easily detectable in the photometric channels of Ariel which cover visible and near-infrared wavelengths (top). A flare of this magnitude could also be detected in the native-resolution spectroscopic light curves at longer wavelengths (bottom).}
    \label{fig:arielrad}
\end{figure}

The question of whether stellar flares can be detected in the infrared is particularly relevant for missions such as Ariel. Stellar flares are traditionally modeled as $\sim9000$\,K black bodies with constant temperature and variable emitting area, meaning that most of their radiative output falls in the near-UV and visible wavelength range. As a result, their infrared contribution is expected to be much weaker, though potentially still detectable. This raises the possibility that flares could represent a source of astrophysical noise for Ariel's primary science goals. Recent observations with James Webb Space Telescope have demonstrated that flares from stars such as TRAPPIST-1 are indeed observable in the near-infrared regime \citep{2023ApJ...959...64H}, with some flare events even seen at 15 $\mu$m \citep{gillon_trap_pc}.

To assess whether such events could be detected by Ariel, we carried out initial simulations using the time-resolved \code{RADYN} flare grid for flares typical to cool dwarf stars. We propagated synthetic flaring time-series spectra through \code{ArielRad} \citep{2020ExA....50..303M}. In the following, we will use GJ367, a bright M-dwarf with an ultrashort-period 0.7 R$_\oplus$ planet \citep{lam_gj367} as an example. To obtain the synthetic spectra, we modeled the quiescent stellar spectrum using a BT-Settl atmosphere model \citep{2014IAUS..299..271A} with $T_\mathrm{eff}=3500$\,K, $\log g = 4.5$, and $[\mathrm{Fe/H}]=0$, 
and added flare spectra from \cite{2024ApJ...969..121K} computed with a one-dimensional radiative transfer model. Specifically, we adopted the \code{mF13-200-3} flare model, which exhibits the characteristic rapid rise and slower decay of stellar flares, together with an additional peak during the rise phase. The flare contribution was scaled to produce a 10\% photometric amplitude in the TESS band, representative of typical M-dwarf flare amplitudes (see Fig.~10 of \citealt{flarecatalog}). To determine the scaling, we convolved the flare spectra with the TESS transmission function and integrated them to construct a flare-only light curve, which was then normalized relative to the integrated stellar spectrum. Figure~\ref{fig:btsettl} shows how the noiseless spectrum evolves, with colors indicating different phases during the flare. Consistent with expectations, most of the flare emission appears at ultraviolet wavelengths, with only a comparatively small excess extending into the infrared.



{Next, we used these synthetic spectra of a flare on GJ367  to illustrate the potential impact of flaring on Ariels observations with realistic noise properties.} 
We simulated the photometric data (VISPhot, FGS1, FGS2) at 10 Hz as this is the standard readout frequency of the FGS. As the NIRSpec channel is located on the same detector as FGS2, we use the same cadence for these data. For the AIRS channels, we set the exposure time to be the point at which 80\% of the well-depth is reached in AIRS Ch0 which, for this target, is 1.46 s. To each time-series, we add random scatter based on the uncertainties predicted by \code{ArielRad}. These initial tests using the time-resolved \code{RADYN} flare grid on an M-dwarf model suggest that stronger flares could be detected using Ariel, as seen in Fig.~\ref{fig:arielrad}. In the photometric channels, the strength of the modeled flare varies significantly but each channel offers sufficient SNR and time-granularity to thoroughly sample the flare. Additionally, based upon this test, the increase in flux in the spectral channels could reach 3\%, which is significant compared 
the transit depth of GJ367\,b which is $\sim$0.02\%. In this case, the duration of the transit is around 40 minutes, comparable with typical flare time scale of $\sim 30$ minutes. However, if the flare is not corrected for it could bias the light curve fitting and lead to biased inferences about the nature of the planet.

Models of stellar flares need to be improved as well---
despite the rapid growth of observational datasets, major limitations remain: flare parameters (e.g., energies) are often estimated from single-band photometry using simplified assumptions that can introduce systematic errors. The standard approach in stellar flare research assumes the optical continuum emission originating from a region with a constant temperature of $\sim$9000~K and a time-variable emitting area, approximated by a blackbody spectrum. 
However, recent multi-band observations of flares on nearby main-sequence stars challenge this paradigm  and indicate that the single-hot-blackbody assumption may not fully describe flare emission \citep[e.g.,][]{Kowalski2019ApJ...871..167K,biczObservationalSignsLimited2025}. 
Flux jumps appear across different wavelength ranges \citep{Kowalski_2013, kowalskiDWARFFLARECONTINUUM2016}, while flare temperatures can increase dramatically, in some cases exceeding $30\,000$~K, even when the emitting area remains nearly constant throughout most of the event. These findings contradict the widely used constant-temperature/variable-area model, suggesting that flare energy estimates derived from it may be significantly biased and that existing theoretical models require stronger observational constraints.

Future work, therefore, will be needed involving the refinement of these models using more realistic flare parameters and time-resolved data, as well as investigating methods to correct for possible flare contamination in Ariel's core science sample.

\section{Summary}
We have investigated the prospects for detecting stellar flares with Ariel and their potential influence on exoplanet spectroscopy. Cross-matching the Ariel Mission Candidate Sample with the TESS flare catalog indicates that most candidate hosts have low observed flare rates and are therefore unlikely to flare during a typical Ariel visit, although the probability becomes substantial for the most active systems. These rates are not completeness-corrected and should be regarded as lower limits above the TESS detection threshold. Our proof-of-concept \code{RADYN} and \code{ArielRad} simulation further shows that a strong flare on a cool dwarf can be readily detected in Ariel's photometric channels and can produce percent-level infrared variations capable of overwhelming planetary atmospheric signals. Ariel's simultaneous VISPhot, FGS, NIRSpec, and AIRS observations will therefore be valuable both for identifying and characterizing flares and for constraining their chromatic contamination of transit and eclipse measurements. Together with evidence that a cadence of a few seconds preserves the principal morphology of complex optical flares, these results establish stellar flares as both a promising ancillary science opportunity and an astrophysical systematic that should be incorporated into Ariel observing and data-analysis strategies.

\section*{Acknowledgements}

We acknowledge the support of the Hungarian National Research, Development and Innovation Office (NKFIH) Élvonal grant KKP 143986.
On behalf of the \textit{"Looking for stellar CMEs on different wavelengths"} project we are grateful for the possibility of using HUN-REN Cloud \citep{MTACloud} which helped us achieve the results published in this paper.

\section*{Data Availability}

The data underlying this study are based on publicly available light
curves from TESS, which can be accessed through the MAST archive
(\url{https://mast.stsci.edu}).
The Ariel Mission Candidate Sample is available online at
\url{https://github.com/arielmission-space/Mission_Candidate_Sample}. The TESS flare catalog of \cite{flarecatalog} is available at \url{
https://zenodo.org/records/14179313}.
All additional data products generated during this study are available
from the corresponding author upon reasonable request.


\bibliographystyle{rasti}
\bibliography{flares}



\bsp	
\label{lastpage}
\end{document}